\documentstyle[11pt]{article}

\begin{document}

\title{Dual Quantum Monte Carlo Algorithm for Hardcore Bosons}

\author{G. G. Batrouni, H. Mabilat\\
Institut Non-Lineaire de Nice, Universit\'e de Nice-Sophia
Antipolis,\\
1361 route des Lucioles, 06560 Valbonne, France}

\maketitle
\begin{abstract}
  We derive the exact dual representation of the bosonic Hubbard model
  which takes the form of conserved current loops. The hardcore limit,
  which corresponds to the quantum spin-${1\over 2}$ Heisenberg
  antiferromagnet, is also obtained. In this limit, the dual partition
  function takes a particularly simple form which is very amenable to
  numerical simulations. In addition to the usual quantities that we
  can measure (energy, density-density correlation function and
  superfluid density) we can with this new algorithm measure
  efficiently the order parameter correlation function, $\langle a_i
  a_j^{\dagger}\rangle , |i-j|\ge 1$. We demonstrate this with
  numerical tests in one dimension.
\end{abstract}

\bigskip
\bigskip

\section{Introduction}

The development of ever more efficient and sophisticated Quantum Monte
Carlo (QMC) algorithms has greatly advanced our understanding of
quantum many body systems and the various phases that exist in such
models (superfluid, Mott insulators, Bose glasses, supersolids etc)
and the transitions between them. Examples of such QMC algorithms
include the ``path integral algorithm'' which was used to study the
details of the Helium superfluid transition in 2 and 3
dimensions\cite{roy}, the World Line algorithm\cite{worldline}, which
was used to study the various phases of the bosonic Hubbard
model.\cite{batrouni1} Improvements of the World Line algorithm for
hard core bosons (infinitely repulsive contact interaction) came in
the form of {\it cluster} algorithms\cite{cluster} where one updates
many variables at a time. Such algorithms converge much faster and
suffer much less from critical slowing down. The next improvement was
to eliminate Trotter errors\cite{trotter} which come from discretizing
the imaginary time ({\it i.e.} inverse temperature) direction. These
continuous imaginary time cluster algorithms\cite{cont-cluster} are
the state of the art in hardcore boson simulations although their
efficiency goes down in the presence of disorder and longer range
interactions like next near neighbour.

In the case of extreme soft-core high density bosonic models a
different algorithm was developed based on duality tranformation. In
this case, the amplitude of the order parameter is assumed constant,
leaving only the phase, giving rise to a model of the $XY$ variety
called the Quantum Phase Model (QPM). The dual of this model is easily
obtained when the Villain aproximation\cite{villain} of the action is
taken, leaving a model of interacting conserved integer current loops
that can be quite easily simulated.\cite{loop} This loop algorithm was
used productively by several groups to study the phases of the QPM
with or without disorder.\cite{loop}

One major disadvantage of the above algorithms is their inability to
measure the correlation function of the order parameter, $\langle
a_ia_j^{\dagger}\rangle$ when $|i-j|$ is greater than $1$. This
quantity is very interesting since in phase transitions the behaviour
of the order parameter is of prime importance.

We will outline below how to perform the duality transformation
exactly for the soft and hard core bosonic Hubbard models. Then,
concentrating on the hardcore case, we will construct the (exact) loop
algorithm and demonstrate how it can be used to obtain the correlation
function of the order parameter in addition to the the usual
quantities of interest (energy, density-density correlation function,
superfluid density).

\section{The Bosonic Hubbard Model}

We are interested in simulating models whose Hamiltonian has the form
\smallskip
$$ H=-t\sum_{\langle
  ij\rangle}(a_{i}^{\dagger}a_{j}+a_{j}^{\dagger}a_{i})-
\sum_{i}\mu_{i}\hat n_{i}+V_{0}\sum_{i}\hat
n_{i}^{2}+V_{1}\sum_{\langle ij \rangle}\hat n_{i}\hat n_{j}+
V_{2}\sum_{\langle\langle ik\rangle\rangle}\hat n_{i}\hat
n_{k}.\eqno(1)
$$ 
\smallskip 
\noindent In this equation, $t$ is the transfer integral (the hopping
parameter) which sets the energy scale (in the simulations we set it
equal to $1$), $V_0$, $V_1$, and $V_2$ are respectively the contact,
the nearest neighbor, and the next near neighbor interactions, which
we always take to be repulsive. On the square lattice, the next near
neighbor is chosen along the diagonal, while in one dimension the
choice is the obvious one.  $a_i$ and $a_i^{\dagger}$ are destruction
and creation operators on site $i$ satisfying the usual softcore boson
commutation laws, $[a_i,a_j^{\dagger}]=\delta_{i,j}$, and $\hat
n_i=a_i^{\dagger}a_i^{}$ is the number operator on site $i$.  In
Eq. (1), $\langle ij\rangle$ and $\langle\langle ik\rangle\rangle$
label sums over near and next near neighbors. The first term which
describes the hopping of the bosons among near neighbor sites gives
the kinetic energy. $\mu_{i}=\mu+\delta_{i}$ is the site dependent
chemical potential. For a clean system {\it i.e.} with no disorder, we
take $\delta_{i}=0$. To include the effect of disorder, we take
$\delta_{i}$ to be a uniformly distributed random number
$-\Delta\le\delta_i\le\Delta$. We have chosen to disorder the system
with a random site energy, which preferentially attracts or repels
bosons to particular sites. Other choices are possible, such as
disordering the hopping parameter, $t$, or some of the interaction
strengths, $V_{0,1,2}$. It is thought that these different ways of
introducing disorder yield similar results.

The quantum statistical mechanics of this system is given by the
partition function Z,
\smallskip
$$ Z = {\rm Tr}e^{-\beta H},\eqno(2)$$
\smallskip 
\noindent where $\beta$ is the inverse temperature. The expectation
value of a quantum operator is given by
\smallskip
$$ \langle {\cal O} \rangle = {1 \over Z} {\rm Tr}({\cal O}e^{-\beta
H}).\eqno(3)$$
\smallskip

\section{Coherent State Representation}

We cannot do numerical simulations directly on this formulation of the
partition function because it is in terms of operators. We must first
find a c-number representation which can be incorporated into a
simulation algorithm. There are several such representations. For
example the wavefunction representation leads to what is known as the
``path integral'' method\cite{roy} and the occupation state
representation leads to the World Line algorithm.\cite{worldline} Here
we will use the {\it coherent states} representation, {\it i.e.}
eigenstates of the destruction operator,
\smallskip
$$a_i|\{ \Phi \} \rangle = \phi(i)|\{ \Phi \} \rangle,\eqno(4a)$$
\smallskip
$$\langle \{ \Phi \} | a_i^{\dagger} = \langle
\{ \Phi \} | \phi^{\star}(i),\eqno(4b)$$
\smallskip
\noindent where the eigenvalues $\phi(i)$ are complex numbers that
are defined on the sites, $i$, of the lattice. In terms of the more
familiar occupation number representation vacuum ($a|0\rangle=0,
a^{\dagger}|0\rangle=|1\rangle$) the coherent state $|{\Phi}\rangle$
is defined as follows
\smallskip
$$|\{\Phi \}\rangle = exp\Biggl (\sum_i(-{{\phi^{\star}(i)\phi(i)}
\over 2} +\phi(i)a_i^{\dagger})\Biggr ) |0\rangle.\eqno(5)$$
\smallskip
\noindent With this normalization we obtain the inner product of two
coherent states
\smallskip
$$\langle \{ \Psi \} | \{ \Phi \} \rangle = exp\Biggl (\sum_i\Bigl
(\psi^{\star}(i)\phi(i)-{1\over 2}\phi^{\star}(i)\phi(i)- {1\over
2}\psi^{\star}(i)\psi(i) \Bigr ) \Biggr ),\eqno(6)$$
\smallskip
\noindent and the resolution of unity
\smallskip
$$1 = \int \prod_i {{d^2\phi(i)} \over {2\pi}} | \{ \Phi \} \rangle \langle \{
\Phi \}|.\eqno(7)$$

Now we write the partition function, Eq. (2), as
\smallskip
$$ Z = {\rm tr} \bigl ( e^{-\delta H} e^{-\delta H} 
e^{-\delta H} ... e^{-\delta H}\bigr ),\eqno(8)
$$
\smallskip

\noindent where $\delta \equiv \beta/L_{\tau}$, $L_{\tau}$ is a large
enough integer such that $\delta \ll 1$. We express the partition
function this way because we can now express the exponentials in
Eq. (8) in a form suitable for easy (albeit approximate)
evaluation. Between each pair of exponentials introduce the resolution
of unity, Eq. 7. Using standard manipulations\cite{negele} we find
\smallskip
$$Z=\int \prod_{r,\tau} {{d^2 \phi(r,\tau)}\over \pi}
e^{-S(\phi^{\star},\phi)},\eqno(9)$$
\smallskip 
\noindent where the action is given to first order in $\delta$
by\cite{caution}
\smallskip
$$S = \sum_{r,\tau} \phi^{\star}(r,\tau)\Delta_{-\tau}\phi(r,\tau) +
\delta \sum_{\tau} H[\phi^{\star}(r,\tau+1),\phi(r,\tau)].\eqno(10)$$
\smallskip
In Eq. 10, $\tau$ denotes imaginary time, {\it i.e.} the inverse
temperature direction $\beta$. Also,
$H[\phi^{\star}(r,\tau),\phi(r,\tau-1)]$ simply means that at the
imaginary time $\tau$, we replace in the Hamiltonian, Eq. 1, the
destruction operator $a_r$ by the complex field $\phi(r,\tau)$, and
the creation operator, $a^{\dagger}_r$, by
$\phi^{*}(r,\tau+1)$.\cite{negele} In this article our notation for
the forward and backward finite difference operators are
\smallskip
$$\Delta_{\mu} \phi(r) \equiv \phi(r+{\hat \mu})-\phi(r),\eqno(11a)$$
\smallskip
$$\Delta_{-\mu} \phi(r) \equiv \phi(r)-\phi(r-{\hat \mu}),\eqno(11b)$$
\smallskip

\noindent It is well known\cite{negele} that in the first term of
Eq. 10 we must have $\Delta_{-\tau}$ and not $\Delta_{\tau}$ although
in the continuum limit one might argue that they both lead to the same
result. Such arguments are incorrect. What is perhaps less strongly
emphasized is that the following ``approximation'',
\smallskip
$$\delta \sum_{\tau} H[\phi^{\star}(r,\tau),\phi(r,\tau-1)]=\delta
\sum_{\tau} H[\phi^{\star}(r,\tau),\phi(r,\tau)] + {\cal O}(\delta
\beta).\eqno(12)$$
\smallskip
\noindent which is often used in the literature, is not correct. The
terms ignored are in fact of order $\beta$, not $\delta \beta$, which
in effect changes the Hamiltonian under consideration.\cite{batrouni2}
This can be easily illustrated using with the quantum harmonic
oscillator.  We will not use this approximation in what follows.

\section{The Duality Transformation}

In what follows we will take $V_0 = V_1 = V_2 = 0$ in order to
simplify the presentation. Nonzero values of these parameters can be
dealt with in a straight-forward way.\cite{batrouni2} Interestingly,
the hardcore boson case (with no near and next near neighbor
repulsion) is obtained directly from this case of zero interaction
(see below).

To effect the duality transformation on the partition function, Eqs. 9
and 10, we follow the method of reference~\cite{batrouni3}. To this end
we first write the coherent states field in terms of the amplitudes
and phases: $\phi(r,\tau)= \alpha(r,\tau)
e^{i\theta(r,\tau)}$. Consequently, 
\smallskip
$$\int \prod_{r,\tau} d^2 \phi(r,\tau) \rightarrow
\int \prod_{r,\tau} \alpha(r,\tau) d\alpha(r,\tau)
d\theta(r,\tau),\eqno(13)$$ 
\smallskip
\noindent where we are dropping irrelevant overall constant
factors. The action \vfil\eject
\smallskip
$$
S = \sum_{ {\vec r}, \tau } \Biggl ( \phi^{\star}({\vec r},
\tau)\Delta_{-\tau}\phi({\vec r},
\tau)~~~~~~~~~~~~~~~~~~~~~~~~~~~~~~~~~~~~~~~~~~~~ 
$$
$$~~~~~~~~~~~~~~~~~~~~~~~-t\delta \sum_{k=1}^{d}\Bigl (
\phi^{\star}({\vec r}, \tau+1)\phi({\vec r}+{\hat k}, \tau)+\\
\phi^{\star}({\vec r}+{\hat k}, \tau+1)\phi({\vec r}, \tau)\Bigr )
\Biggr ),\eqno(14)
$$
\smallskip
\noindent becomes
\smallskip
$$
S = \sum_{{\vec r}, \tau} \Biggl ( \alpha({\vec
r},\tau)e^{-i\theta({\vec r},\tau)} \Bigl ( \alpha({\vec
r},\tau)e^{i\theta({\vec r},\tau)}-\alpha({\vec
r},\tau-1)e^{i\theta({\vec r},\tau-1)} \Bigr
)
$$
$$~~~~~~~~~~~~~-t\delta\sum_{k=1}^{d}\Bigl ( \alpha({\vec r},
\tau+1)\alpha({\vec r}+{\hat k},\tau)e^{i(\theta({\vec r}+{\hat
k},\tau)-\theta({\vec r},\tau+1))}
$$
$$~~~~~~~~~~~~~~~~~~~~~~~~~+\alpha({\vec r}, \tau)\alpha({\vec
r}+{\hat k},\tau+1)e^{i(\theta({\vec r},\tau)-\theta({\vec
r}+{\hat k},\tau+1))}\Bigr ) \Biggr ).\eqno(15)
$$
\smallskip
\noindent Here we took the model to be in $d$ space dimensions
indicated by the index $k=1,...,d$. ${\hat k}$ is a unit vector in the
the $k$th direction.

We see in this equation that the phase, $\theta({\vec r}, \tau)$,
which is a site variable, appears only as differences of near neighbor
sites\cite{footnote}. Thus we can see the beginings of an $XY$-like
model. The fact that we always have this combination of variables
motivates us to change the variable of integration from the site
variable $\theta({\vec r}, \tau)$ to the {\it link} or {\it bond}
variables $\theta_k({\vec r}, \tau)\equiv \Delta_k \theta({\vec r},
\tau)$ and $\theta_{\tau}({\vec r}, \tau)\equiv \Delta_{\tau}
\theta({\vec r},\tau)$. In two space dimensions, the partition
function thus becomes\cite{batrouni3}
\smallskip
$$
Z = \int \prod_{r,\tau,\mu=1,2,3}\Bigl ( \alpha(r,\tau) d\alpha(r,\tau)
d\theta_{\mu}(r,\tau) \Bigr ) ~~~~~~~~~~~~~~~~~~~~~~~~~~~~~~~~~~~
$$
$$~~~~~~~~~~~~~~\prod_{plaquettes} \Biggl (\delta \Big [
e^{i\epsilon_{\mu \nu \rho}\Delta_{\nu}\theta_{\rho}({\vec r},\tau)}-1\Bigr
] \Biggr )~~~~~~~~~~~~~~~~~~~~~~~~~~~~~~~~~~~~~~~~~~
$$
$$~~~~~~~~~~~~~\prod_{\vec r} \Biggl ( \delta \Bigl [
e^{i\sum_{\tau}\theta_{\tau}({\vec r},\tau)}-1\Bigr ]\Biggr )
e^{-S},~~~~~~~~~~~~~~~~~~~~~~~~~~~~~~~~~\eqno(16) 
$$
\smallskip
\noindent where $\epsilon_{\mu \nu \rho}$ is the totally antisymmetric
tensor in three dimensions. Note that the product over plaquettes is
over all space-space and space-time plaquettes. The $\delta$-functions
are the only Jacobian of this variable change.\cite{batrouni3} Its
geometrical interpretation is simple. Even though the model is now
expressed in terms of bonds, it is actually a site model. It is clear
from the definitions of the link variables that if we sum them along
any directed closed path the result is zero (mod $2\pi$). This is
known as the Bianchi identity which is lost when the model is
expressed in terms of bonds and needs to be enforced as a
constraint. The first set of $\delta$-functions in the above equation
enforces the ``local'' Bianchi identities, {\it i.e.}  those due to
topologically trivial loops. The second set enforces the ``global''
Bianchi identities, {\it i.e.} those due to topologically nontrivial
loops in the imaginary time direction due to the periodic boundary
conditions. These constraints have several interesting relationships
to various geometrical aspects of the theory.\cite{batrouni3} Here we
will simply exploit its relationship to the duality transformation. As
was shown in Ref.~\cite{batrouni3} the dual variables are the Fourier
conjugates to these constraints. In other words, the Fourier
expansions of these $\delta$-functions
\smallskip
$$
\prod_{plaquettes} \Biggl (\delta \Big [ e^{i\epsilon_{\mu
\nu \rho}\Delta_{\nu}\theta_{\rho}({\vec r},\tau)}-1\Bigr ] \Biggr )=
\sum_{\{ l_{\mu}({\vec r},\tau)=-\infty\} }^{+\infty} e^{i\sum_{{\vec
r},\tau}l_{\mu}({\vec r},\tau)\epsilon_{\mu
\nu \rho}\Delta_{\nu}\theta_{\rho}({\vec r},\tau)}
$$
$$
~~~~~~~~~~~~~~~~~~~~~~~~~~~~~~~~~~~~~~~~~~~~~~~~~~~=\sum_{\{ l({\vec
r},\tau)=-\infty\} }^{+\infty} e^{-i\sum_{{\vec
r},\tau}\theta_{\rho}({\vec r},\tau)\epsilon_{\mu
\nu \rho}\Delta_{-\nu}l_{\mu}({\vec r},\tau)},\eqno(17)
$$
\smallskip
\noindent and
\smallskip
$$
\prod_{\vec r} \Biggl ( \delta \Bigl [
e^{i\sum_{\tau}\theta_{\tau}({\vec r},\tau)}-1\Bigr ]\Biggr )=
\sum_{\{ n_{\tau}({\vec r})=-\infty \} }^{+\infty} e^{i\sum_{{\vec
r},\tau}n_{\tau}({\vec r})\theta_{\tau}({\vec r},\tau)},\eqno(18)
$$
\smallskip
\noindent 
immediately give the dual variables. In the case of the local Bianchi
constraints, Eq 17, the dual is the integer valued bond variable
$l_{\mu}({\vec r},\tau)$, while in the global case, Eq 18, the dual
variable is the integer valued field $n_{\tau}({\vec r})$. Note that
whereas $l_{\mu}({\vec r},\tau)$ is a {\it vector} bond variable which
depends on both coordinates ${\vec r}$ and $\tau$ and which has
components in the $x, y, {\rm and} \tau$ directions, the variable
$n_{\tau}({\vec r})$ is global and only points in the time
direction. It depends only on the spatial coordinates and gives the
value of the current flowing in the time direction from $\tau=0$ to
$\tau=L_{\tau}$ where $L_{\tau}$ is the number of steps in the
imaginary time direction.\cite{batrouni3} Also note that the form in
which the dual variable $l_{\mu}({\vec r},\tau)$ appears is always
$\epsilon_{\mu \nu \rho}\Delta_{-\nu}l_{\mu}({\vec r},\tau)$. So we
define the local electric current
\smallskip
$$
j_{\rho}({\vec r},\tau) = -\epsilon_{\mu \nu
\rho}\Delta_{-\nu}l_{\mu}({\vec r},\tau).\eqno(19)
$$
\smallskip
\noindent Due to the totally antisymmetric tensor $\epsilon$, it is
clear that the local integer current $j_{\rho}({\vec r},\tau)$ is
conserved.

Substituting Eqs 17, 18 and 19 in Eq 16 we can integrate over the
original variables, $\theta_{\mu}({\vec r},\tau)$ and $\alpha({\vec
r},\tau)$. The details of the integration will be given
elsewhere.\cite{batrouni2} This leaves the partition function
expressed only in terms of the dual variables $n_{\tau}, j_{\mu}$ and
$s_k$. The new variable $s_k$ is positive semidefinite integer valued
and is nothing but the dual to the amplitude field $\alpha$. The
partition function thus becomes:
\smallskip
$$ 
Z = \sum_{\{ n_{\tau},j_{\mu},s_{k}\} } (t\delta)^{\sum_{{\vec
r},\tau}2s_{k}({\vec r}, \tau)} \Biggl (\prod_{{\vec r},\tau}
{{[n_{\tau}({\vec r})+j_{\tau}({\vec r},\tau)]!}\over {[n_{\tau}({\vec
r})+j_{\tau}({\vec r},\tau)-M({\vec r},\tau)]!}}\Biggr)~~~~~~~~~~~
$$
$$~~~~~~~~~~~~~~~~~~~~~~~~~\Biggl (\prod_{{\vec r},\tau} [1+\delta \mu({\vec
r})]^{n_{\tau}({\vec r})+j_{\tau}({\vec r},\tau)-M({\vec r},\tau)}
\Biggr )$$
$$~~~~~~~~~~~~~~~~~~~~~~~~~\Biggl ( \prod_{{\vec r},\tau,k}{1\over {[s_k({\vec
r},\tau)]! [s_k({\vec r},\tau)+j_k({\vec r},\tau)]!}}\Biggr ),\eqno(20)
$$
\smallskip
\noindent where 
\smallskip
$$ M({\vec r},\tau) \equiv \sum_{k=1,2} \bigl (s_k({\vec
r},\tau)+s_k({\vec r}-{\hat k},\tau)+j_k({\vec r}-{\hat
k},\tau).\eqno(21) $$
\smallskip
\noindent In this expression, which will be greatly simplified below,
we allowed for the possibility of disorder in the chemical
potential. Also, even though originally the values of $n_{\tau}$ ran
from $-\infty$ to $+\infty$ (Eq. 18), the $\theta_{\tau}$ integrals
impose the condition that $n_{\tau}$ is positive semidefinite. In fact
we can easily prove\cite{batrouni2} that the total current traversing
a bond in the time direction is nothing but the number of bosons
traversing that bond.  In addition, it is understood that all the
arguments of factorials are positive semidefinite. This imposes
several severe constraints on the allowed configurations which we will
exploit to simplify the partition function.

For simplicity, although this is not necessary, we will consider the
no disorder case, {\it i.e.} $\mu(r)\rightarrow \mu$. In addition, we
will consider the very important case of hardcore bosons where
$n_{\tau}({\vec r})+j_{\tau}({\vec r},\tau)$ can take only the values
$0$ or $1$. Combining this with the previously mentioned constraints
on the arguments of the factorials, allows us to solve for the allowed
electric loop configurations.\cite{batrouni2} In this case the
partition function simplifies drastically. For example for the one
dimensional model it becomes
\smallskip
$$
Z_{Q} = \sum_{\{ n_{\tau}=0,1\} }\sum_{\{ j_{\mu}=0,\pm 1\} } e^{\beta
\mu\sum_{x}n_{\tau}(x)} \bigl (t\delta\bigr
)^{\sum_{x,\tau}|j_x(x,\tau)|},\eqno(22)
$$
\smallskip
\noindent for the grand canonical ensemble and
\smallskip
$$ Z = \sum_{\{ j_{\mu}=0,\pm 1\} } \bigl (t\delta\bigr
)^{\sum_{x,\tau}|j_x(x,\tau)|},\eqno(23) $$ 
\smallskip
\noindent for the canonical case. These are the duals to the hardcore
boson Hubbard model. The interpretation is very simple: The dual is a
model of conserved integer current loops that take the values $0$ or
$1$ in the time direction and $0,\pm1$ in the space directions. The
partition function is a sum over all deformations with each spatial
hop (in this case $x$) costing $t\delta$. This is a very simple and
appealing form and is very amenable to numerical simulations.

The algorithm is now very simple. For example for the canonical case,
we start with all local currents zero, and with the desired number of
nonzero global time currents, $n_{\tau}$, corresponding to the number
of bosons in the system. Then we visit each plaquette and randomly
choose to add to it a positive or negative elementary current loop
(plaquette). This attempt is rejected if (i) it introduces negative
time currents, or (ii) it introduces currents larger than $1$. If this
test is satisfied, we accept this current loop in accordance with
detailed balance: If $(t\delta)^2$ is greater than or equal to a
uniformly distributed random number we accept the change.

\section{Physical Quantities}

Now that we have the partition function and the algorithm we need
expressions for the physical quantities we want to measure. The
expression for the energy can be easily obtained from $<E>=-
{{\partial}\over {\partial \beta}}{\rm ln}Z$. This gives
\smallskip
$$
\langle E \rangle = -{1\over \beta} \langle \sum_{x,\tau}
|j_x(x,\tau)| \rangle.\eqno(24)
$$
\smallskip
\noindent The numerical values for $\langle E\rangle$ obtained with
the above algorithm and Eq 24 are shown in Fig. 1 and compared with
the results of exact diagonalization. We see that the agreement
between the two results is excellent.

The expression for the equal time density-density correlation function
is
\smallskip
$$
\langle n(x_1)n(x_2) \rangle = \langle \Bigl ( n_{\tau}(x_1) +
j_{\tau}(x_1,\tau) \Bigr )\Bigl ( n_{\tau}(x_2) + j_{\tau}(x_2,\tau)
\Bigr )\rangle,\eqno(25)
$$
\smallskip
\noindent since the number of particles on site $(x,\tau)$ is given
by $( n_{\tau}(x) + j_{\tau}(x,\tau))$. We compare the numerical
values of this correlation function with the exact values in
Fig. 2. Again we see that agreement is excellent. 

The superfluid density is, in general, related to the winding number,
$W$, of current configurations by\cite{batrouni1}
\smallskip
$$
\rho_s = {{\langle W^2 \rangle }\over {2t\beta}}.\eqno(26)
$$
\smallskip
\noindent We have so far only discussed local moves in our Monte Carlo
algorithm, and such moves can never change the winding number of a
configuration. Therefore, the system is stuck in the winding number
sector of the initial configuration. If the initial configuration has
$W=0$, this will give $\langle W^2\rangle =0$ and therefore zero
superfluid density. This obstacle can be overcome with a trick. Define
the {\it total} $x$-current at a given time $\tau$ by
$J_x(\tau)=\sum_x j_x(x,\tau)$ and calculate the Fourier transform of
the current-current correlation function
\smallskip
$$ 
{\tilde{\cal J}}(\omega) = \sum_{\tau, \tau_0} e^{i{{2\pi}\over
L_{\tau}}\omega \tau} \langle J_x(\tau_0)J_x(\tau_0 +
\tau)\rangle.\eqno(27)
$$
\smallskip
\noindent We can show\cite{batrouni1} that ${\tilde{\cal J}}(\omega
\rightarrow 0)= W^2$ which allows us to calculate the superfluid
density $\rho_s$. In Fig. 3 we show $\rho_s$ as a function
$(\rho_c-\rho)$ where $\rho_c=1$ is the critical density at which the
hardcore boson system becomes an incompressible Mott insulator and
$\rho$ is the boson density.  From scaling arguments\cite{fisher} we
expect $\rho_s \sim (1-\rho)^{\nu z}$ with $\nu z=1$.\cite{footnote2}
Our numerically obtained value on a small system is $\nu z=0.96$ in
excellent agreement with the theoretical values.

These three quantities can be easily measured by existing QMC
algorithms mentioned earlier. What is qualitatively new here is that
we can measure the correlation function of the order parameter very
easily. Let ${\cal N}(r)$ be the number of jumps of length $r$ in a
spatial direction (in this case $x$) in a given loop configuration. We
can show\cite{batrouni2}
\smallskip
$$
\langle a(x_1) a^{\dagger}(x_2) \rangle = {1\over {\beta
\delta^{|x_1-x_2|-1}}} \langle \sum_{l=0}^{{{L_x}\over 2}-|x_1-x_2|}
(l+1){\cal N}(|x_1-x_2|+l)\rangle,\eqno(28)
$$
\smallskip
\noindent where $|x_1-x_2|\ge 1$. This quantity is shown in Fig. 4 for
$\beta=0.5$ and $4$ for $L_x=8$. We see that agreement with exact
values is excellent for $\beta=0.5$ and gets worse for $\beta=4$ as
$|x_1-x_2|$ increases. The reason is that the exact diagonalization
results sum over all winding number sectors whereas our simulation
stays in the $W=0$ sector. At high temperature ($\beta=0.5$) the
correlation length is short and the boundary conditions are less
important so we get good agreement. At lower temperatures the
correlation length is longer and consequently the boundary conditions
become very important for small systems. For the more interesting
larger system sizes the boundary conditions will become less
important.

\section{Conclusions}

We have outlined how to perform the duality transformation exactly for
the soft and hard core bosonic Hubbard models. In the hard core case
the dual model is particularly simple. The dual partition function was
used to construct a loop Quantum Monte Carlo algorithm and we showed
that the numerical results agree very well with those of exact
diagonalization for a system with $8$ sites and $4$ bosons.

In particular we showed that with our algorithm we can calculate
easily the order parameter correlation function, $\langle a(x_1)
a^{\dagger}(x_2)\rangle$ which is very difficult to do with most other
QMC algorithms. This interesting quantity is very important for the
elaboration of the quantum phase transitions exhibited by the system.

This representation of the Hubbard model has many common features with
the World Line representation for which very efficient cluster
algorithms have been constructed.\cite{cluster,cont-cluster} It,
therefore, seems possible to improve the efficiency of this algorithm
in the same way. A cluster algorithm would improve the convergence
properties and in addition would lift the restriction to zero winding
number sector.\cite{cluster} This would greatly improve the
calculation of the order parameter correlation function. This is
curently under investigation.

{\it Note added.} After the completion of this work we became aware of
two other algorithms which allow to calculate the correlation function
of the order parameters. These are the ``worm'' algorithm\cite{worm}
and the cluster algorithm with a new interpretation of the
clusters.\cite{brower} 

\section{Acknowledgements}

We thank Richard Scalettar for many very helpful discussions and for
the exact diagonalization results used in this article.

\vfill\eject

\centerline{\bf FIGURE CAPTIONS}

\begin{itemize}
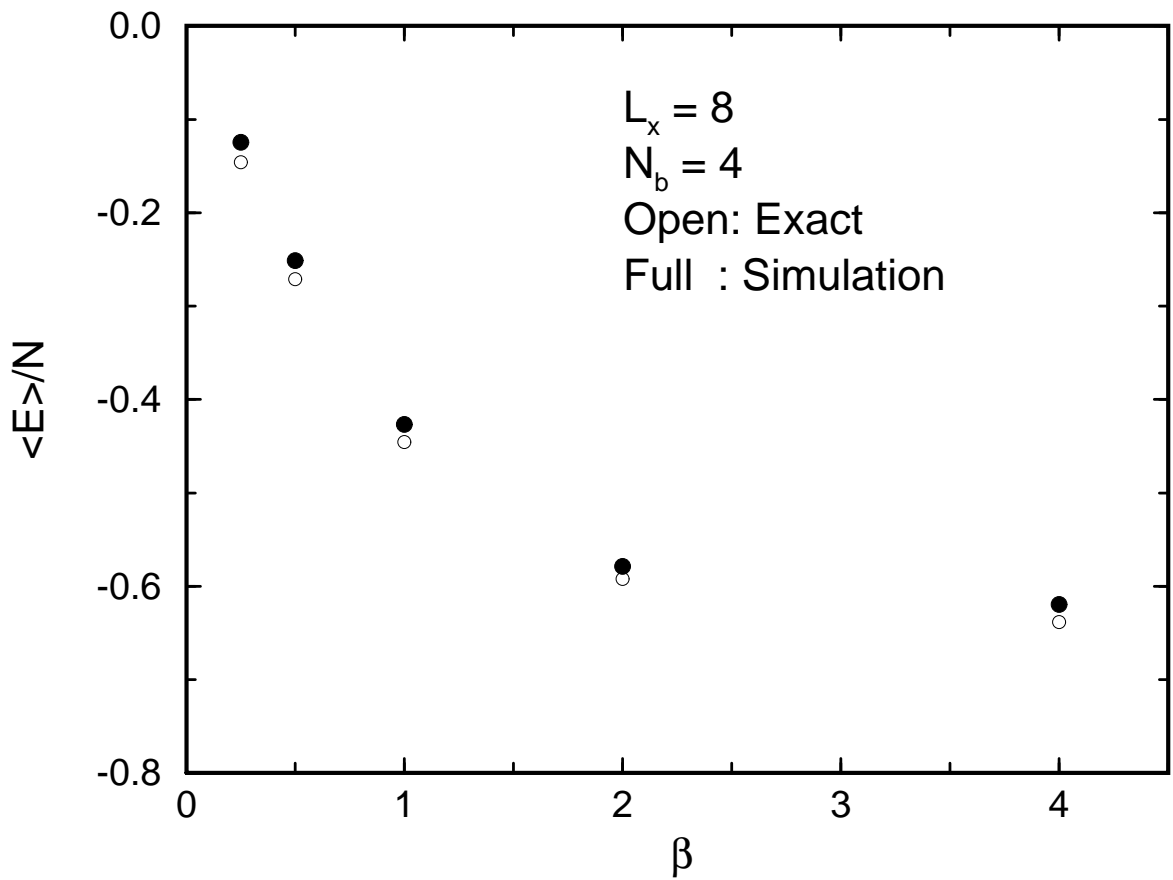
\item[FIG. 1] The average energy per site as a function of
       $\beta$. $L_x, N_b$ are the number of lattice sites and bosons
       respectively.
  
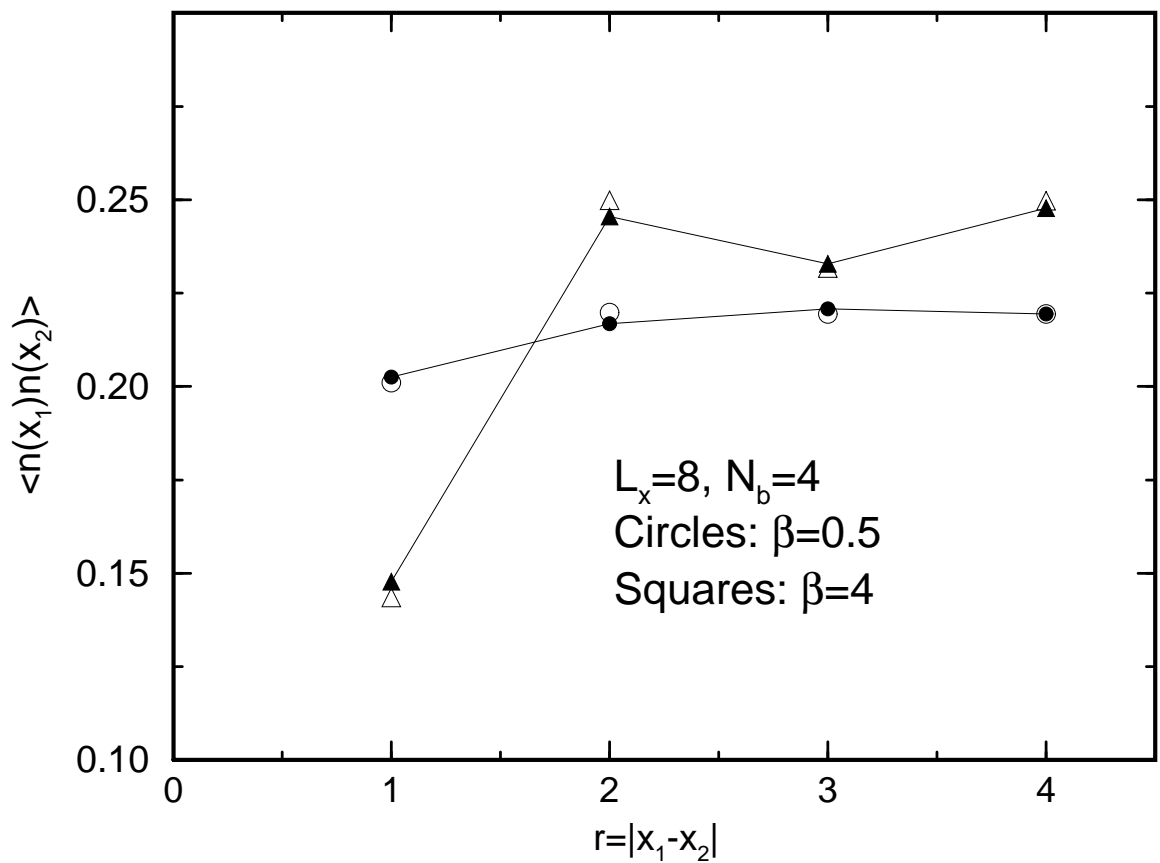
\item[FIG. 2] The density-density correlation function as a function
       of distance. Open symbols are exact diagonalization values,
       full symbols are simulation results. The lines are just to
       guide the eyes.

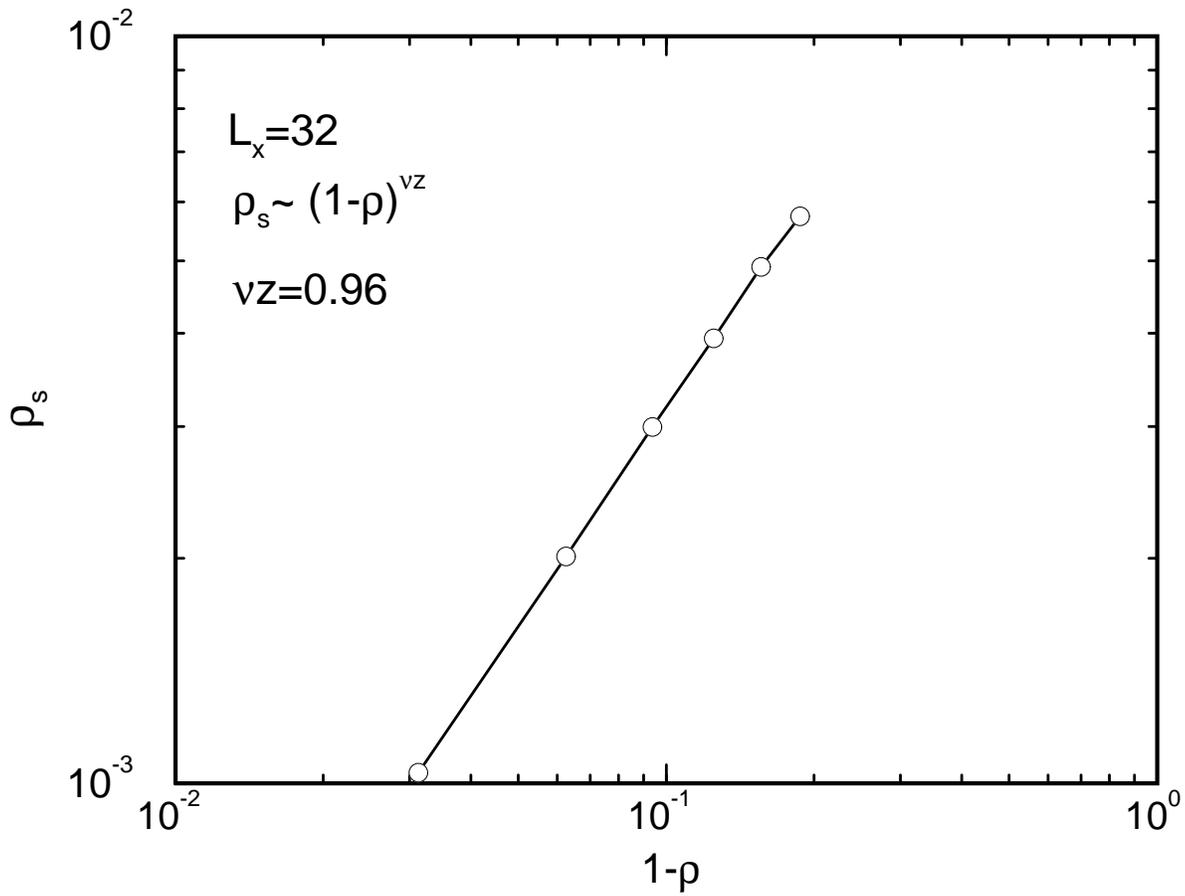
\item[FIG. 3] The superfluid density, $\rho_s$, as a function of
       $(1-\rho)$. The theoretical value of $\nu z$ is 1.

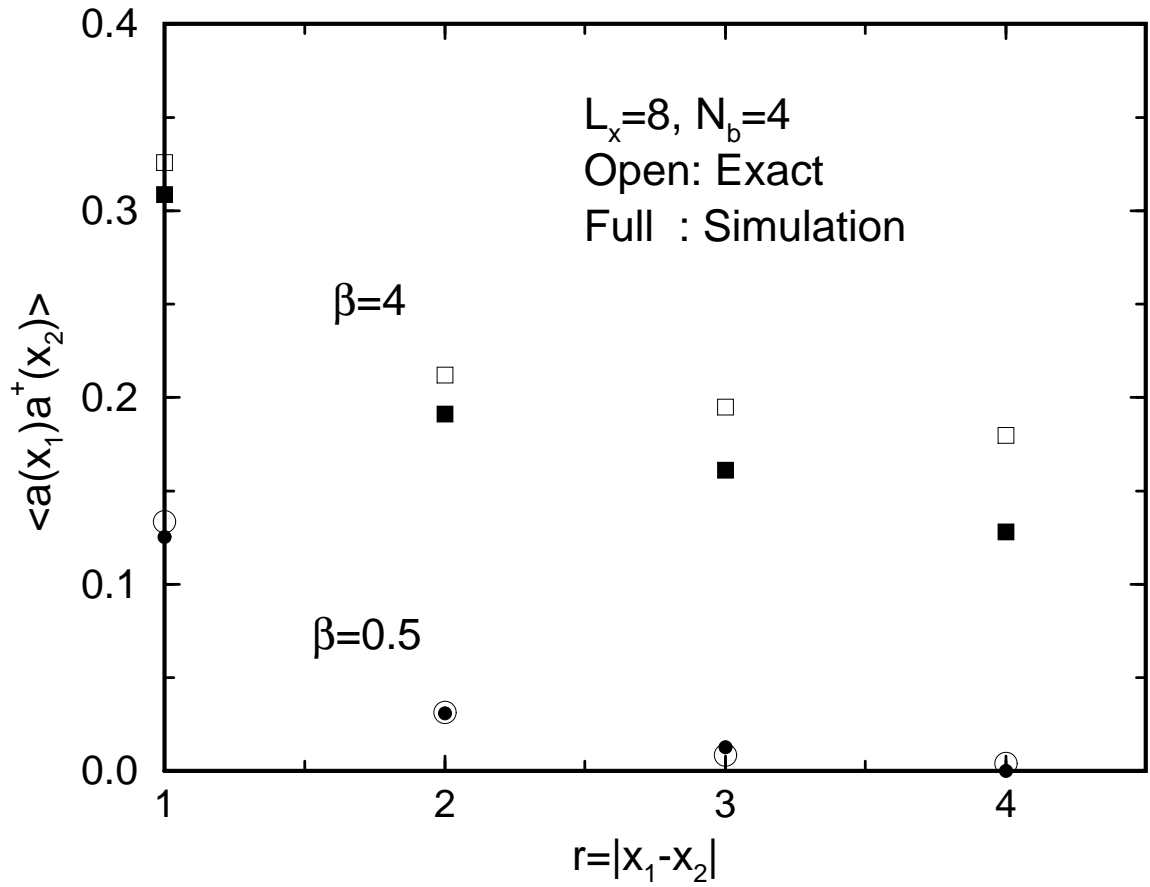
\item[FIG. 4] The correlation function of the order parameter, $<a(x_1)
       a^{\dagger}(x_2)>$ as a function of distance for two values of
       $\beta$.

\end{itemize}

\vfill\eject


\begin{thebibliography}{99}
\bibitem{caution} This is not quite correct. If we normal order
$e^{-\delta H}$ only to order $\delta$ we will end up with a dual
model that does not allow hops of more than one lattice spacing at a
time. There is no physical basis for this restriction, and in addition
it leads to underestimates of the kinetic energy. The correct
procedure is to normal order the exponential of the kinetic energy
operator to all orders in $\delta$. This is easy since this operator
is quadratic,
\bibitem{footnote} Differences of next near neighbors can always be
                expressed as the sums of differences of near
                neighbors.
\bibitem{footnote2} $\nu$ is the correlation length critical exponent
and $z$ is the dynamic critical exponent. They are expected to have
the mean-field values $\nu=1/2, z=2$.
\end{thebibliography}
\end{document}